\documentclass[twocolumn,amsmath,amssymb,prl]{revtex4-1}                    

\usepackage{titlesec}
\usepackage[colorlinks,bookmarks=false,citecolor=blue,linkcolor=red,urlcolor=blue]{hyperref}
\usepackage{epsfig}
\usepackage{color}
\usepackage{subfigure}
\usepackage{graphicx}    
\usepackage{dcolumn}     
\usepackage{lipsum}      
\input{insbox}
\usepackage[T1]{fontenc}

\newcommand{\be}{\begin{equation}}
\newcommand{\ee}{\end{equation}}

\definecolor{drkgr}{rgb}{0.05,0.6,0.2}

\begin{document}

\title{Crystal-field effects competing with spin-orbit interactions in NaCeO$_2$}

\author{P.~Bhattacharyya}
\affiliation{Institute for Theoretical Solid State Physics, Leibniz IFW Dresden, Helmholtzstr.~20, 01069 Dresden, Germany}

\author{U.~K.~R{\"o}{\ss}ler}
\affiliation{Institute for Theoretical Solid State Physics, Leibniz IFW Dresden, Helmholtzstr.~20, 01069 Dresden, Germany}

\author{L.~Hozoi}
\affiliation{Institute for Theoretical Solid State Physics, Leibniz IFW Dresden, Helmholtzstr.~20, 01069 Dresden, Germany}

\begin{abstract}
\noindent
Ce compounds feature a remarkable diversity of electronic properties, which motivated extensive
investigations over the last decades.
Inelastic neutron scattering represents an important tool for understanding their underlying electronic
structures but in certain cases a straightforward interpretation of the measured spectra is hampered
by the presence of strong vibronic couplings.
The latter may give rise to extra spectral features, which complicates the mapping of experimental data 
onto standard multiplet diagrams. 
To benchmark the performance of embedded-cluster quantum chemical computational schemes for the case 
of $4f$ systems, we here
address the Ce 4$f^1$ multiplet structure of NaCeO$_2$, an antiferromagnet with $D_{2d}$ magnetic-site
symmetry for which neutron scattering measurements indicate only weak vibronic effects. 
Very good agreement with the experimental results is found in the computations, which
validates our computational approach and confirms NaCeO$_2$ as a 4$f$ magnet in the
intermediate coupling regime with equally strong 4$f$-shell spin-orbit and crystal-field interactions.
\end{abstract}

\date\today
\maketitle

{\it Introduction.\,}
Spin-orbit interactions received in recent years enormous attention.
New insights and new ideas have led to new physical models, new concepts, and new research paths,
as for example the Kitaev honeycomb model \cite{KITAEV2006,Khal_2009} and related extensive
investigations.
A very interesting aspect is the role of crystalline fields in realizing a particular type of
magnetic ground state.
For octahedral ligand coordination of $d$- or $f$-metal magnetic centers, not only the octahedral
crystal-field splittings are relevant but also additional splittings related to lower-symmetry fields
--- tetragonal, trigonal etc.
The latter may originate from distortions of the ligand cages away from a regular, cubic octahedron
and/or peculiarities of the crystalline lattice, e.\,g., from having a layered structure.
The precise nature and strength of the underlying ligand/crystal fields are relevant to properties
and parameters such as magnetic moments, single-ion anisotropies, intersite exchange couplings.
Spin-orbit interactions and crystal-field splittings meet sometimes on the same energy scale, in
both $d$- \cite{CaIrO3_marco_2014,CaIrO3_niko_2012,RuCl3_gael_2018} and $f$-electron systems
\cite{Ce_garnets_seijo_2014,KCeO2_qc_2020,KCeO2_bordelon_2021,NaCeO2_bordelon_2021,Pr213_calder_2021}.
This may give rise to e.\,g. unexpected magnetic ground states \cite{RuCl3_gael_2018}.
It may also hamper a straightforward interpretation of experimental data \cite{Ce_garnets_seijo_2014}
and may pose problems to computational modelling \footnote{See, e.\,g., the discussion in \cite{
Ce_garnets_seijo_2014}}.
The Ce$^{3+}$ oxide compounds are in this context representative, with spin-orbit couplings (SOCs) and
crystal-field splittings of similar magnitude \cite{Ce_garnets_seijo_2014,KCeO2_qc_2020,
KCeO2_bordelon_2021,NaCeO2_bordelon_2021}.
The 4$f$-shell multiplet structure is relatively simple in these systems: 
for a Ce$^{3+}$ 4$f^1$ ion in symmetry lower than $O_h$, seven Kramers doublets are expected.
The lowest three and upper four are degenerate in the case of a free ion, defining $J\!=\!5/2$ and
$J\!=\!7/2$ free-ion terms.
Interestingly, in delafossite \cite{KCeS2_gael_2020,KCeO2_bordelon_2021} and pyrochlore \cite{
Ce227_gaudet_2019,Ce227_sibille_2020} structures with $D_{3d}$ Ce-site symmetry, one extra excitation
is observed experimentally in the lower energy range (i.\,e., within the $J\!=\!5/2$\,-like energy
window), likely arising from strong vibronic couplings.
For benchmark {\it ab initio} multiplet calculations in the intermediate coupling regime with equally
strong 4$f$-shell spin-orbit and crystal-field interactions, we here choose NaCeO$_2$ as test case.
It features $D_{2d}$ Ce-site symmetry and complications as seen in $D_{3d}$ setting \cite{KCeS2_gael_2020,
KCeO2_bordelon_2021,Ce227_gaudet_2019,Ce227_sibille_2020,CsYbSe2_lawrie_2022} do not arise in this
geommetry \cite{NaCeO2_bordelon_2021}.
Our study adds useful reference data to investigations addressing the electronic structure and the
physical properties of Ce oxide compounds, providing a solid basis for extensions towards the computation
of total energy landscapes and vibronic excitations.

\begin{figure}
\includegraphics[width=1\columnwidth]{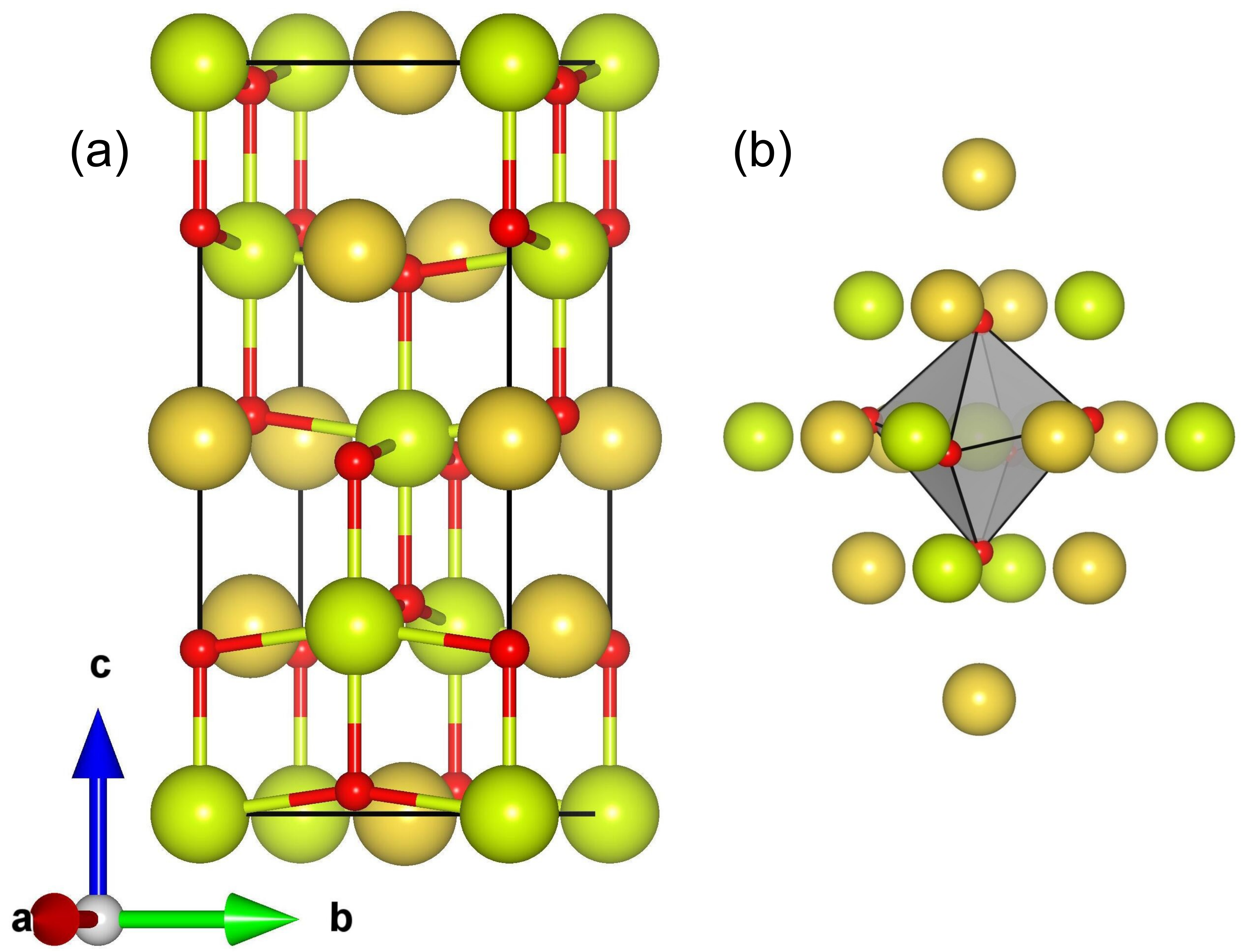}
\caption{
(a) Unit cell of NaCeO$_2$, plot using the {\sc vesta} visualization program \cite{vesta}.
Green, yellow, and red spheres indicate Ce, Na, and O species, respectively.
(b) 25-site cluster considered in the quantum chemical calculations.
The crystalline environment (not depicted) was modeled as a large array of point charges in the computations.
}
\label{fig_1}
\end{figure}

{\it Computational details.\,}
Tetragonally distorted, edge-sharing CeO$_6$ octahedra form a bipartite diamond magnetic lattice in
NaCeO$_2$, as depicted in Fig.~\ref{fig_1}(a).
To understand the specificities of the Ce$^{3+}$ 4$f^1$ multiplet structure in this particular
crystallographic setting, we carried out detailed quantum chemical embedded-cluster calculations. 
For this purpose, not only a CeO$_6$ octahedron was considered at the quantum mechanical level but
also the eight Ce and ten Na nearest neighbors, see Fig.~\ref{fig_1}(b).
The crystalline environment of this 25-site unit was modeled as a large array of point charges that
reproduces the crystalline Madelung field within the cluster volume.
To generate this collection of point charges we employed the {\sc ewald} package \cite{Klintenberg_et_al,
Derenzo_et_al}. 
In a first set of computations, a fully ionic picture with formal valence states was
assumed: Ce$^{3+}$, O$^{2-}$, and Na$^+$.

The actual quantum chemical calculations were performed using the {\sc molpro} suite of programs
\cite{Molpro}.
The numerical investigation was initiated as a complete active space self-consistent field (CASSCF)
computation \cite{MCSCF_Molpro,olsen_bible} with all seven 4$f$ orbitals of the central Ce site incorporated in
the active orbital space.
The seven crystal-field states associated with the 4$f^1$ manifold were obtained from a state-averaged
\cite{olsen_bible} variational optimization.
Ce 4$f$ and O 2$p$ electrons on the central CeO$_6$ octahedron of the 25-site cluster were
subsequently considered in a multireference configuration-interaction (MRCI) with single and double
excitations (MRSDCI) \cite{MRCI_Molpro,olsen_bible}.
Finally, spin-orbit calculations were carried out in terms of both CASSCF and MRSDCI states \cite{Berning_et_al}.
We used energy-consistent quasirelativistic pseudopotentials \cite{Dolg_Stoll_Preuss} and Gaussian-type
valence basis sets of quadruple-$\zeta$ quality \cite{Cao_Dolg} for the central Ce ion, whereas for
O ligands of the central CeO$_6$ octahedron we employed all-electron correlation-consistent 
polarized basis sets of triple-$\zeta$ quality \cite{Dunning}.
We adopted large-core pseudopotentials including the 4$f$ subshell in the core for the eight Ce
nearest neighbors \cite{Dolg1989}.
Large-core pseudopotentials were also considered for the ten adjacent Na cations \cite{Fuentealba_Na}.

{\it Ce 4$f^1$ mutliplet structure.\,}
The Ce 4$f^1$ valence configuration is associated with fascinating properties, ranging from 
low-dimensional frustrated magnetism and possibly a spin-liquid ground state in
KCeSe$_2$ \cite{KCeSe2_sefat_2022} to Kondo physics in various intermetallic compounds \cite{Fulde}
and multipolar states in Ce hexaboride \cite{CeB6_inosov_2016,CeB6_thalmeier_2021}
and Ce oxide pyrochlores \cite{Ce227_sibille_2020}. 
CASSCF and MRSDCI results for the Ce 4$f^1$ electronic structure in NaCeO$_2$, both without and with
SOC, are presented in Table \ref{tab:excited_energy}.
We here employed crystallographic data as reported in Ref.~\cite{NaCeO2_bordelon_2021}.
NaCeO$_2$ displays a $I4_{1}/amd$ tetragonal lattice \cite{LiYbO2_bordelon_2021} (see Fig.\;\ref{fig_1});
the Wyckoff positions of Ce, Na, and O are 4a (0,\,0,\,0), 4b (0,\,0,\,1/2) and 8e(0,\,0,\,0.21921),
respectively, while the experimentally determined unit-cell parameters are $a$=$b$=4.77860 and c=11.04277 
\AA \ \cite{NaCeO2_bordelon_2021}.
A CeO$_6$ octahedron features two distinct types of Ce-O links.
The Ce-O bond lengths, 2.41 and 2.42 \AA , are not very different; what plays a more important role in
splitting the set of six ligands into two symmetry inequivalent groups is the farther-neighbor linkage,
see Fig.\;\ref{fig_1}.
There are also three different O-Ce-O bond angles, of 81.9, 91.1, and 98.1 degrees.
As the Ce-site point group symmetry is $D_{2d}$, the $f$ levels are split into three nondegenerate
$a_1$, $a_2$, and $b_2$ and two sets of doubly degenerate $e$ crystal-field sublevels.

As far as the crystal-field levels in NaCeO$_2$ are concerned, very large splittings of up to 250
meV are computed  (first columns in Table\;\ref{tab:excited_energy}, results without including SOC),
a few times larger than the strength of the spin-orbit coupling constant \cite{4f_atanasov_16,
KCeO2_qc_2020}.
This is then reflected in the splittings among the spin-orbit Kramers doublets (last columns in Table
\ref{tab:excited_energy}, SOC included).
Experimental estimates for the relative energies of the lowest two excited states are available from
inelastic neutron scattering (INS) measurements \cite{NaCeO2_bordelon_2021}.
Very good agreement is found between peak positions in the INS spectra and MRSDCI+SOC results for
those two low-lying on-site excited states: 118 vs 121 and 125 vs 126 meV, respectively.
The {\it ab initio} quantum chemical data also show that the two lowest excited states lie at about
the same distance with respect to the ground state and to the next excited Kramers doublet. 
All these electronic-structure peculiarities indicate that in NaCeO$_2$ the $J\!=\!5/2$ and $J\!=\!7/2$
nomenclature is not appropriate.

\begin{table}[!b]
\caption{ 
Ce$^{3+}$ 4$f^1$ multiplet structure in NaCeO$_2$, relative energies in meV.
Notations as in $D_{2d}$ symmetry are used for the crystal-field (SOC not included) and spin-orbit
states (+\,SOC).
}
\begin{tabular}{l c c | c c l}
\hline
\hline

           &CASSCF  &MRSDCI &MRSDCI+SOC &INS            &\\

\hline

$^2\!A_1$  &0       &0      &0          &0$\pm$5        &$\Gamma_6$\\
$^2\!E$    &88      &92     &121        &117.8$\pm$1.8  &$\Gamma_7$\\
           &88      &92     &126        &124.8$\pm$1.7  &$\Gamma_7$\\
$^2\!A_2$  &110     &110    &249        &--             &$\Gamma_6$\\
$^2\!E$    &234     &238    &369        &--             &$\Gamma_6$\\
           &234     &238    &370        &--             &$\Gamma_7$\\
$^2\!B_2$  &253     &252    &437        &--             &$\Gamma_7$\\
\hline
\hline
\end{tabular}
\label{tab:excited_energy}
\end{table}

Splittings as large as here were earlier found in 4$f^1$ delafossites \cite{KCeO2_qc_2020,KCeO2_bordelon_2021},
honeycombs \cite{Pr213_calder_2021}, and garnets \cite{Ce_garnets_seijo_2014}.
Intriguingly, one extra excitation is experimentally observed in delafossites \cite{KCeS2_gael_2020,
KCeO2_bordelon_2021}, presumably related to vibronic effects \cite{CeF3_gerlinger_1986}.
This provides strong motivation for more systematic {\it ab initio} quantum chemical investigations
in 4$f^1$ compounds, delafossites but also other variaties such as the NaCeO$_2$ 
system addressed here.
With anomalies in the measured spectra (e.\,g., extra peaks as in delafossites  \cite{KCeS2_gael_2020,
KCeO2_bordelon_2021} and pyrochlores \cite{Ce227_gaudet_2019,Ce227_sibille_2020}), a clear assignment of 
the excitations is problematic at the model-Hamiltonian level.

Not surprisingly, the MRSDCI treatment brings only minor corrections to the CASSCF excitation energies,
as seen by comparing data in the first two columns of Table\;\ref{tab:excited_energy};
having only one 4$f$ electron for the leading ground-state configuration, there are only weak on-site
4$f$--semicore and intersite Ce\,4$f$\,--\,O\,2$p$ correlations showing up post-CASSCF.
But substantial MRSDCI corrections were found in previous studies for larger filling of the 4$f$ shell,
in e.\,g. 4$f^{13}$ compounds \cite{Zangeneh_et_al}.

\begin{table}[!t]
\caption{ 
Ground-state $g$ factors in NaCeO$_2$, by CASSCF+SOC, MRSDCI+SOC, and INS \cite{NaCeO2_bordelon_2021}
\footnote{On the basis of model-Hamiltonian fits of INS peak positions.}.
}
\begin{tabular}{l c c c l}
\hline
\hline

              &$g_{ab}$  &$g_c$    \\

\hline

CASSCF+SOC    &1.12      &0.71     \\
MRSDCI+SOC    &1.11      &0.66     \\
INS           &1.41      &1.00     \\
\hline
\hline
\end{tabular}
\label{tab:g_factors}
\end{table}

Knowing that valence-semicore and ligand-metal charge-transfer-type correlation effects are not
significant, the very good agreement between computational and experimental results convincingly
validates the material model employed here: 
`central' quantum mechanical cluster, buffer region consisting of large-core effective potentials
and less sophisticated valence basis functions, plus point-charge embedding.
As concerns the latter, we checked how the $f$-$f$ excitation energies depend on the precise
values chosen for the ionic charges of the extended lattice, i.\,e., reduced the formal $+3$ (Ce)
and $-2$ (O) to $+2.6$ (Ce) and $-1.8$ (O)
\footnote{In other words, the Madelung potential in the cluster region (25 sites) corresponds in this
new set of calculations to a periodic structure with $+1$, $+2.6$, and $-1.8$ point charges at the
Na, Ce, and O lattice sites. Overall charge neutrality is preserved by subsequent slight readjustment
of O charges in the immediate neighborhood of the 25-site central fragment.}.
The $f$-$f$ excitation energies computed this way are essentially the same as in Table\;\ref{tab:excited_energy} \footnote{Variations within 1\;meV are found.},
which shows that
although less elaborated than e.\,g. embeddings constructed on the basis of prior Hartree-Fock
\cite{HF_embeddings} or density-functional \cite{DFT_embeddings_1,DFT_embeddings_2} periodic
calculations, a point-charge representation of the extended crystalline surroundings is effective
for ionic materials.
A more sensitive aspect is how cation species in the immediate vicinity of the central quantum
unit are modeled:
using for the nearby cations [see Fig.\,1(b)] just bare positive charges may lead to spurious
orbital polarization at the boundaries of the quantum mechanical region \cite{PC_embeddings_3,
PC_embeddings_5,Hozoi2011,Bogdanov_2011}.

Based on the spin-orbit MRSDCI and CASSCF wave functions, we also calculated Ce-ion $g$ factors,
using the Gerloch-McMeeking formula \cite{g-factor} and following the procedure outlined in Ref.
\cite{Bogdanov_et_al};
results are presented in Table\;\ref{tab:g_factors}, along with experimentally measured $g$ factors.
It is seen that the $g$ factors are fairly anisotropic.
The magnitude of the $g$ factors is somewhat on the lower side in the calculations as compared to the
experimental estimates.

{\it Conclusions.\,}
In sum, the accuracy of an embedded-cluster material model relying on point-charge embedding and a small
buffer region between the point-charge array and the quantum mechanically modelled cluster is verified
for the case of a 4$f^1$ oxide, NaCeO$_2$. 
The system is well suited to this purpose since accurate experimental data are available for the on-site
$f$-$f$ excitation energies and dynamical correlation effects \cite{olsen_bible} are modest for the
4$f^1$ configuration.
The latter feature, in particular, eliminates one possible source of errors in the electronic-structure
calculations.
Very good agreement with experimental results is found in the quantum chemical computations, as also seen
in the case of $d$-electron systems with one particle per site \cite{Bogdanov_2011}.
This validates the type of embedding scheme employed here.
Our analysis also indicates large 4$f$-shell crystal-field splittings of up to 250 meV, which renders
the $J$-multiplet nomenclature inappropriate \cite{J_terms_magnani_05} and confirms NaCeO$_2$ \cite{
NaCeO2_bordelon_2021} as a 4$f$ magnet in the intermediate coupling regime with equally strong 4$f$-shell
spin-orbit and crystal-field interactions.

 \

{\it Acknowledgments.\,}
We thank T.~Petersen, M.~S.~Eldeeb, M.~M.~Bordelon, and S.~D.~Wilson for discussions, U.~Nitzsche for technical
assistance, and the German Research Foundation (project number 441216021) for financial support.

\bibliography{ff_naceo2_mar07}

\end{document}